\def\comment#1{}

\documentstyle[pra,aps,epsf]{revtex}

\title{Quantum Langevin equation from forward-backward path integral}
\author{H. Kleinert and S.V. Shabanov\footnote{Alexander
von Humboldt fellow,
on leave from Laboratory
of Theoretical Physics, JINR, P.O.Box 79, Moscow, Russia
\\~\\
\sf Freie Universit\"at Berlin preprint FUB--HEP/95--1\\
}}
\address{Institut f\"{u}r Theoretische Physik,
        Freie Universit\"{a}t Berlin,
        Arnimallee 14, D-14195 Berlin}
\date{\today}

\begin{document}
\maketitle

\begin{abstract}
The quantum Langevin equation is derived from the Feynman-Vernon
forward--backward
path integral  for
a density matrix of a quantum system in a thermal
oscillator  bath.
We exhibit the mechanism by which the
classical, $c$-valued
noise in the Feynman-Vernon theory
turns into an
operator-valued quantum noise.
The quantum noise fulfils  a
characteristic commutation relation
which ensures the unitarity of the time evolution
in the quantum Langevin equation.
\end{abstract}

\def\qle{quantum Langevin equation}
\def\ad{{\rm ad\,}}

\section{Introduction}
The {\em quantum Langevin equation\/}
\begin{equation}
m\ddot {\hat x}_t+ \gamma\dot{\hat x}_t+V'(\hat x_t)= \hat \eta_t
  \label{ql}
\end{equation}
describes successfully the temporal
behavior of a quantum mechanical
point-like particle
with the action
\begin{equation}
{\cal A}^S[x]= \int\limits_{0}^{t} dt'
\left[\frac m2 {\dot x_t^2} -V(x_t)\right]\
\label{sysac}
\end{equation}
in the presence of dissipation
\cite{senitzky,lax}.
The time argument is indicated by a subscript, for notational brevity.
The dissipation is accounted for in Eq.~(\ref{ql})
by the two phenomenological terms:
the friction term $\gamma\dot {\hat x}_t$, and
the quantum noise operator
$ \hat \eta_t$ which satisfies the commutation rules
\begin{eqnarray}
[\hat{\eta}_t,\hat{\eta}_{t'}] &= &2i\hbar\gamma\partial_t\delta(t-t')\
\label{qn1}\end{eqnarray}
and has the expectation values
\begin{eqnarray}
\frac 12\langle[\hat{\eta}_t,\hat{\eta}_{t'}]_+\rangle_{\hat{\eta}} &= &
K_T(t,t') \ ,
\label{qn2}\end{eqnarray}
with
\begin{equation}
K_T(t,t')=\frac{\gamma \hbar}{\pi}\int\limits_{0}^{\infty}
d\omega\omega\coth\frac{\hbar\omega}{2kT}~
\cos\omega(t-t').
\label{4}\end{equation}
The prefactor $ \gamma $ is once more the friction
constant, in accordance with the fluctuation-dissipation theorem.
As usual,
 $T$ denotes the temperature and $k$ the Boltzmann constant.
Remarkably, the temporal evolution governed by the quantum Langevin equation
preserves the canonical commutation relation, i.e.,
quantizing
 $[\hat x_t, \hat p_t] =i \hbar$ with $\hat p_t = m\dot{\hat x}_t$
 at one time, this remains true at all times
\cite{gardiner}.

The quantum Langevin equation (\ref{ql})
has been derived \cite{ullersma,kac,mazur}
from the Heisenberg equations of motion of
the quantum system
coupled to a thermal environment
consisting of a bath of infinitely many harmonic oscillators.
The coupling and the spectral distribution of the oscillators
are chosen in such a way that
 the Heisenberg equations for the environment
coordinates can be solved explicitly,
and that
the standard friction term arises in (\ref{ql}).
The general procedure is described
in \cite{zwanzig,ford}.

The \qle{} can be used to
derive {\em Kubo's
stochastic Liouville equation\/}
\cite{kubo}
for the temporal evolution of the
density matrix of the system \cite{gardiner}.
Solutions of this equation
are in agreement with
experimental data.

The same  Kubo equation can be obtained
  \cite{leggett}
from a completely different
description of the quantum system due to Feynman and Vernon \cite{vernon,fey}.
Here the density matrix of the quantum system in
an
oscillator bath is represented by a
path integral which
contains a
fluctuating noise variable $ \eta _t$.
In contrast to the
\qle, the fluctuating noise in this formulation is classical.

In ordinary quantum mechanical path integrals, there exists a simple way
of going from fluctuating classical variables to Heisenberg operators.
For the classical fluctuating noise variable in the
Feynman-Vernon path integral, on the other hand,
the relation to the quantum noise operator $\hat  \eta  _t$
is unknown.
The purpose of this paper is to exhibit this relation and to derive the
\qle{} from the Feynman-Vernon path integral.

\section{The forward--backward path integral
for the density matrix of a system in a thermal bath}

Consider the quantum mechanical system
described by the action (\ref{sysac}) in contact with an oscillator bath.
The density operator $\hat{\rho}_t$
of the total system obeys the Neumann equation
\begin{equation}
i\hbar\partial_t \hat{\rho_t}=[\hat{H},\hat{\rho}_t]\ ,
\label{neq}\end{equation}
where the Hamilton operator is the sum of three terms
\begin{equation}
\hat{H}=\hat{H}^S+\hat{H}^B+ \hat{H}^I               .
 \label{ham1}\end{equation}
The first term
$\hat{H}^{S}$
describes the quantum system by itself,
$\hat{H}^{B}$ the oscillators in the heat bath, and
$\hat{H}^{I}$ the interaction of the system with the bath.
A formal solution to Eq. (\ref{neq}) reads
\begin{equation}
\hat{\rho}_t=\exp\left(-\frac{i\hat{H}t}{\hbar}\right)\hat{\rho}_0
\exp\left(\frac{i\hat{H}t}{\hbar}\right)\ .
\end{equation}

Let
$X$
denote collectively an
infinite set of
oscillator coordinates in the heat bath.
Then a matrix element of $\hat{\rho}_t$ has
the time evolution
\begin{equation}
\langle xX|\hat{\rho}_t|x'X'\rangle = \int d\bar{x}d\bar{x}'
d\bar{X}d\bar{X}'
\langle xX|e^{-i\hat{H}t/\hbar}|\bar x\bar X\rangle
 \langle \bar x\bar X|\hat{\rho}_0|\bar x'\bar X'\rangle
\langle \bar x'\bar X'|e^{i\hat{H}t/\hbar}|x'X'\rangle   .
\label{2.3}\end{equation}
By time-slicing the matrix elements of the
two evolution operators on the right-hand side (one of which
working forward, the other  backward
in time), they can be represented by corresponding path integrals
\cite{fey,vernon,kleinert},
\begin{equation}
\langle xX|e^{-i\hat{H}t/\hbar}|\bar x\bar X\rangle =
\int {\cal D}x{\cal D}X e^{i{\cal A}[x,X]/\hbar}\ ,
\end{equation}
with the boundary conditions $x_0=\bar x,\ x_t=x$ and $X_0=\bar X,\
X_t=X$, and
\begin{equation}
\langle \bar x'\bar X'|e^{i\hat{H}t/\hbar}|x'X'\rangle =
\int {\cal D}x'{\cal D}X' e^{-i{\cal A}[x',X']/\hbar},
\end{equation}
with the boundary conditions $x'_0=\bar x',\ x'_t= x'$ and $X'_0=\bar X',\
X'_t=X'$.
The functional
\begin{equation}
{\cal A} [x, X]
 \equiv {\cal A}^S[x] +  {\cal A}^B [X]
    + {\cal A}^I[x,X]
\label{act}\end{equation}
is the classical action associated with the
Hamiltonian (\ref{ham1})
of the total system.

We are interested
in the temporal behavior of the system
coordinate $x_t$ regardless
of the behavior of the environment variables $X_t$, i.e.,
 we look for the so-called
{\em reduced description\/} of the total quantum system.
This is obtained by tracing out
the environment coordinates in (\ref{2.3}).
The system and the bath may be
assumed
to be decoupled at some initial time, say $t=0$.
At that time, $\hat{\rho}_0=
\hat{\rho}_0^S\hat{\rho}_0^B$,
and we can rewrite (\ref{2.3})  as
\begin{eqnarray}
\rho^S_t(x,x') &\equiv &\int dX \langle xX|\hat{\rho}_t|x'X\rangle
= \int dx_1dx_1'U_t(xx',x_1x_1')\rho^S_0(x_1,x_1')\ ,
\label{2.8}\end{eqnarray}
where
$U_t(xx',x_1x_1')$
has the forward--backwardpath integral representation
 (the hyphen is pronounced {\em minus\/})
\begin{eqnarray}
U_t(xx',x_1x_1') &= & \int {\cal D}x_+{\cal D}x_-\ F[x_+,x_-] \exp\frac{i}{\hbar}
\left\{{\cal A}^S[x_+]-{\cal A}^S[x_-]\right\}
\label{2.7}\end{eqnarray}
with $x_+(0)=x_1, x_+(t)=x, x_-(0)=x'_1, x_-(t)=x'$.
For variables  with subscripts $\pm$ we place the time arguments in parentheses, to avoid a
pileup of subscripts.
The functional
\begin{eqnarray}
 F[x_+,x_-]&=&\int d\bar Xd\bar X' \rho_0^B(\bar X,\bar X')\\
 &\ &\times\ \int dX\int {\cal D} X_+{\cal D} X_-
\exp\frac{i}{\hbar}\left\{{\cal A}^I[x_+, X_+]-{\cal A}^I[x_-, X_-
]
+{\cal A}^B[X_+]-{\cal A}^B[X_-]\right\}
\label{14}\end{eqnarray}
is Feynman's {\em influence functional\/} \cite{fey,vernon,kleinert}.
The forward--backward path integral
over the bath coordinates is taken with the boundary
conditions $X_+(0)=\bar X, X_+(t)=X, X_-(0)=\bar X'$ and $X_-(t)=X$.

The simplest bath action containing
oscillators
of frequency $ \omega $ with coordinates
 $X_ \omega (t)$ is
\begin{eqnarray}
{\cal A}^B[X] &= &\frac 1{2}\int\limits_{0}^{t}dt'\int\limits_{0}^{\infty}
d\omega
\left[\dot{X}^2_\omega - \omega^2X_\omega^2\right]\ .
\label{bath1}
\end{eqnarray}
Assuming an initial equilibrium density operator
for the bath,
\begin{equation}
\hat{\rho}^B_0 = \exp(- \hat{H}^B/kT),
\label{bath4}\end{equation}
the interaction
leading to the correct
\qle{} is  \cite{gardiner,kleinert,schmid}
\begin{equation}
{\cal A}^I[x,X] = (2\gamma/\pi)^{1/2}\int\limits_{0}^{t}dt' x
\int\limits_{0}^{\infty}d\omega \dot{X}_\omega \ .
\label{bath3}
\end{equation}
Since all integrals are Gaussian,
the influence functional  can be
calculated explicitly \cite{fey,vernon,kleinert}. The result is
\begin{eqnarray}
F[x_+,x_-]&=&
F_0\exp\left(\frac{i\gamma}{\hbar}\int\limits_{0}^{t}dt' y\dot{x}\right)
\exp\left[-\frac{1}{2\hbar^2}\int\limits_{0}^{t}
dt'dt'' y_{t'} K_T(t',t'')y_{t''}\right]
\label{infl0} \\
&\equiv&
\exp\left(\frac{i\gamma}{\hbar}\int\limits_{0}^{t}dt' y\dot{x}\right)
F_{\rm fl}[y]\ ,                \label{infl}
\end{eqnarray}
where we have introduced new variables
\begin{equation}
x\equiv (x_++x_-)/2,~~~~y\equiv x_+-x_-.
\label{newv}\end{equation}
The constant $F_0=F[0,0]$ in  (\ref{infl0}) is
the initial value of the influence functional, where
the system and bath are decoupled, i.e., $F_0=
{\rm Tr}\hat{\rho}^B_0=\int dX\rho_0^B(X,X)$; the exponential
in (\ref{infl}) is the dissipation part of the influence functional,
while the functional $F_{\rm fl}[y] $ describes thermal
and quantum fluctuations.
The two factors contribute in completely different ways
to the Langevin equation.
 \comment{The first gives a
friction, the second a noise terms in the
Langevin equation.}

Introducing a new function
$\tilde{\rho}_t^S(x,y)\equiv \rho_t^S(x+y/2,x-y/2)$
and substituting (\ref{infl0}) into (\ref{2.7}) and (\ref{2.8}), we obtain
the evolution equation
\begin{eqnarray}
\tilde{\rho}_t^S(x,y)
&= &\int dx'dy' \tilde{U}_t^S(xy,x'y')\tilde{\rho}_0^S(x',y')\ ,
\label{2.170}\end{eqnarray}
where  $\tilde{U}_t^S(xy,x'y')$ is given by the functional integral
\begin{eqnarray}
\tilde{U}_t^S(xy,x'y') &= &
F_0\int {\cal D}x{\cal D}y\exp i{\cal A}_{T}[x,y]/\hbar\ ,
\label{2.17}\end{eqnarray}
with the
boundary conditions
$x_0=x', x_t=x, y_0=y', y_t=y$.
The  temperature-dependent
action is
\begin{equation}
{\cal A}_{T}[x,y]=\int\limits_{0}^{t}dt' \left[m\dot{y}\dot{x} - \gamma
y\dot{x} - V(x+y/2) + V(x-y/2) + \frac{i}{2\hbar}y\hat{K}_T y\right]\ .
\label{effa}
\end{equation}
For brevity, we have written $\hat{K}_Ty_{t'} $ for the integral
$\int_0^tdt''K_T(t',t'')y_{t''}$ in the last term.
Due to this term, the temperature-dependent action is nonlocal in time.
Note that at $t=0$,
$U_0^S(xy,x'y')$ reduces to $
F_0\delta(x-x')\delta(y-y')$.

The temporal nonlocality of the action (\ref{effa})
can be removed by
means of a random {\em noise\/} variable
with the correlation function
\begin{equation}
\langle \eta_t \eta_{t'} \rangle_  \eta =
K_T(t,t') .
  \label{18.399}
\end{equation}
  We simply make use of the
Gaussian identity
\begin{eqnarray}
\exp\left(-\frac{1}{2\hbar^2}\int\limits_{0}^{t}dt'
y\hat{K}_Ty\right) &=& \int {\cal D}\eta\exp\int\limits_{0}^{t}dt'
\left(-\frac 12\eta\hat{K}_T^{-1}\eta +\frac i\hbar \eta y\right)
\nonumber\\
&\equiv& \langle\exp\frac i\hbar\int\limits_{0}^{t}dt' y\eta\rangle_\eta\ .
\label{25}
\end{eqnarray}
associated with (\ref{18.399}).
Thus, if we agree to average all equations at the end with respect to
the noise variable,
we may replace
the
effective action
in the evolution operator
(\ref{2.17}) by
\begin{equation}
{\cal A}_{ \eta }[x,y]
=\int\limits_{0}^{t}dt' \left[
m\dot{y}\dot{x} - \gamma y\dot{x}
 - V(x+y/2) + V(x-y/2) +  y \eta \right]\ .
\label{effap}\end{equation}

In the limit of high temperatures,
the noise fluctuations
become large and local,
\begin{equation}
 \langle \eta_t \eta_{t'} \rangle_  \eta = K_T(t,t') =
 2kT\gamma \delta (t-t') + {\cal O}(1/T)\ .
\label{lnc}\end{equation}
Equation (\ref{25}) shows that
the average size of the
fluctuating variable $y$
goes to zero like
$\hbar /\sqrt{kT}$.
Hence the action (\ref{effa})
becomes local, the potential difference
$V(x+y/2) - V(x-y/2) $ can be approximated by
$V'(x)y$, and
(\ref{effap})
turns into
\begin{equation}
{\cal A}_{ \eta }[x,y]
=\int\limits_{0}^{t}dt' \left[
- m\ddot{x} - \gamma
\dot{x}
 - V'(x) +  \eta \right]y\ .
\label{effap1}\end{equation}
An integration by parts has been performed
in the first term of
 the integrand in (\ref{effap}),
absorbing a boundary term
 in the prefactor
of the path integral (\ref{2.17}).
 We can now perform the functional integral over $y$ and find
 the classical version of the Langevin equation (\ref{ql}):
\begin{equation}
m\ddot { x}_t+ \gamma\dot{ x}_t+V'( x_t)=  \eta_t\ ,
  \label{cl}
\end{equation}
which describes  classical Brownian motion.
The classical behavior at large $T$ is a
 consequence of $\hbar$ and $T$ appearing  in the combination
$\hbar \omega /2kT$ in ({\ref 4}), so that $T \rightarrow  \infty$
is equivalent to $\hbar \rightarrow 0$.
%

There have been attempts to include quantum effects
into this
classical equation
by simply
replacing the local
noise correlation (\ref{lnc})
by the finite temperature one
(\ref{18.399}), while
maintaining
the linear approximation $V'(x)y$
to $V(x+y/2) - V(x-y/2)$.
The result has been called a {\em quasiclassical Langevin equation\/}
\cite{schmid,qcle}.
Obviously, such an approximation
can be reasonable
 only
for nearly
harmonic potentials (see also the remarks in Ref. \cite{leggett}, p. 589).

It is surprising that
by converting the variables $x_t$ and $ \eta _t$ in
(\ref{cl}) into operators in (\ref{ql}),
{\em all\/} terms in the expansion
of $V(x+y/2) - V(x-y/2)$ in powers of $y$
can be accounted for, as we shall prove.

\section{Kubo's stochastic Liouville equation}

Equations (\ref{2.170})
and
(\ref{2.17}) resemble the evolution equation
of a Schr\"odinger wave function
whose role is now played by the
density function  $\tilde{\rho}_t^S(x,y)$.
This analogy is helpful in deriving a
Schr\"odinger-like differential equation for $\tilde{\rho}_t^S$.
This equation is most easily found by going over
from a
Lagrange-type path integral (\ref{2.17}) to
a Hamiltonian one, expressed in terms
 the position and canonical momentum variables
\cite{cite1}.
The effective action (\ref{effa}) goes over into
\begin{equation}
{\cal A}_{T}[x,y]
\rightarrow {\cal A}_{T}[p_x,p_y,x,y]=\int\limits_{0}^{t}dt'\left(
p_x\dot{x}+ p_y\dot{y}- H_T\right)\ ,
\label{ham} \end{equation}
where
\begin{equation}
 H_T=\frac{1}{m}\left(p_x+\gamma y\right)p_y +V(x+y/2)-V(x-y/2) -
\frac{i}{2\hbar}y\hat{K}_Ty
\label{canh}\end{equation}
plays the role of a Hamiltonian.
For the noise-dependent effective action (\ref{effap})
we find an analogous action
$ {\cal A}_ \eta [p_x,p_y,x,y]$
involving the
 Hamiltonian
\begin{equation}
H_ \eta
=\frac{1}{m}\left(p_x+\gamma y\right)p_y +V(x+y/2)-V(x-y/2) -
 y\eta.
\label{canhn}\end{equation}

Using the effective action
(\ref{ham}),
we obtain the alternative path integral
representation
for the time evolution operator in
(\ref{2.170}):
\begin{eqnarray}
\tilde{U}_t^S(xy,x'y') &= & F_0
\int{\cal D}p_x{\cal D}p_y
{\cal D}x{\cal D}y\exp i{\cal A}_T[p_x,p_y,x,y]/\hbar\ .
\label{2.17hh}\end{eqnarray}
A similar equation holds for a fixed
noise with the action ${\cal A}_ \eta [p_x,p_y,x,y]$.

Since the
action ${\cal A}_ \eta [p_x,p_y,x,y]$ is
local in time, we conclude that
for a fixed noise $ \eta $, the associated noise-dependent ({\em noisy})
density matrix
$\tilde{\rho}^S_t[ \eta ]$
obeys the Schr\"odinger-like equation
\begin{equation}
i\hbar\tilde{\rho}_t^S[\eta] =\hat{H}_\eta\tilde{\rho}_t^S[\eta]\ ,
\label{sla2}\end{equation}
with $ \hat{H}_\eta $ being the
operator arising from (\ref{canhn})
by substituting
 $p_x\rightarrow -i\hbar\partial_x,\
p_y\rightarrow -i\hbar\partial_y$.
The noise average of the solution
to (\ref{sla2}) is the  density matrix of the system:
\begin{equation}
\tilde{\rho}_t^S(x,y) =\langle\tilde{\rho}_t^S[\eta]\rangle_\eta\ .
\label{3.3}\end{equation}
Equation (\ref{sla2}) supplemented by
(\ref{18.399}) and (\ref{3.3}) is called Kubo's
stochastic Liouville equation \cite{kubo},
which we have thus derived from
the Feynman-Vernon forward--backward path integral.

Note that there is no analogous procedure
to obtain
a differential equation for the density matrix $\tilde\rho_t^S(x,y)$
 from the Hamiltonian
(\ref{canh}),  due to the nonlocality of the
 $\hat{K}_T$-term.
Only in the limit of large temperatures, when
 $H_T$
becomes local due to (\ref{lnc}),
there exists a
Schr\"odinger-like equation which is the Fokker-Planck equation
at finite friction.

It must be pointed out that the transition
from the Hamiltonians
$ H_T$ and
$H_\eta$ to their operators
has an ordering
ambiguity in  the term $\gamma yp_y/m$
in $\hat{H}_T$ and $\hat{H}_\eta$
\cite{cite1}.
In the time-sliced path integral
we must decide whether to write
$\gamma y_np_{y\,n}/m$ or
$\gamma y_np_{y\,n-1}/m$ where the subscript $n$
numbers the
time slice.
 Since the ordering
is independent of temperature,
we resolve
the ambiguity
in the limit of large temperatures:
from the well-known
Fokker-Planck equation we determine  the
correct operator ordering
of the term $ \gamma yp_y/m$ to be
$-i\hbar\gamma y\partial_y/m$.

\section{Classical noise versus quantum noise}

The noise in Kubo's equation is a $c$-number, and its relation
to an operator-valued noise of
the quantum Langevin equation (\ref{ql})
has been an outstanding puzzle, as pointed out in the Introduction.
To find this relation, we observe that it is possible
to remove the temporal nonlocality
in the initial effective action (\ref{effa})
just as easily with the help of an operator-valued noise $\hat  \eta $.
This was, in fact, how the Feynman-Vernon path integral was derived in
Ref.~\cite{kleinert} [see Eq.~(18.162)].
An operator-valued noise possessing the properties
(\ref{qn1})--(\ref{qn2}) can be chosen
as a sum of all Heisenberg operators of the
oscillator velocities.
Explicitly:
\begin{eqnarray}
\hat{\eta}_t &=& \sqrt{2\gamma/\pi}\partial_t
\int\limits_{0}^{\infty}d \omega
e^{i\hat{H}^Bt/\hbar}\ \hat{X}_\omega\ e^{-i\hat{H}^Bt/\hbar}
\label{noise} \\
&=& i(\gamma/\pi)^{1/2}\int\limits_{0}^{\infty}d\omega
\sqrt{\omega}\left(e^{i\omega t}\hat{a}^\dagger_\omega -
e^{-i\omega t}\hat{a}_\omega\right)\ ;
\label{3.6}
\end{eqnarray}
where
$a^\dagger_\omega $ and $\hat{a}_\omega$
are time-independent creation and annihilation operators
with the usual commutation rules:
\begin{equation}
[\hat{a}_\omega,\hat{a}_{\omega'}]=0,~~~
[\hat{a}_\omega^\dagger,\hat{a}_{\omega'}^\dagger]=0,
~~~[\hat{a}_\omega,\hat{a}_{\omega'}^\dagger]
=\hbar\delta (\omega-\omega')\ .
\end{equation}

The noise operator (\ref{3.6}) satisfies
the commutation rule (\ref{qn1}).
The correlation function (\ref {qn2})
follows if we define
$\hat  \eta $-averages
as bath averages:
\begin{equation}
\langle[\hat{\eta}_t,\hat{\eta}_{t'}]_+\rangle_{\hat{\eta}} \equiv
{\rm Tr}~(\hat  \rho ^B_0[\hat{\eta}_t,\hat{\eta}_{t'}]_+)\ .
\label{aver0}\end{equation}
Thus the operator (\ref{noise})  has  precisely
 the  properties of
the quantum noise variable in  the quantum Langevin equation (\ref{ql}).

In the operator representation
of the noise variable,
the influence functional
is given by \cite{cite2}
\begin{equation}
F[x_-,x_+]={\rm Tr}\,
\left[\hat\rho_0^B
\hat {T}_C\  \exp \frac{i}{\hbar} \int_C
        dt' x_C \hat\eta_C
\right]\ ,
\label{3.4a}\end{equation}
where
$C$ is a closed-time contour
encircling tightly the interval $[0,t]$
in the complex $t$ plane and  $\hat {T}_C$
is the ordering operator along this contour.
The subscript $C$ distinguishes the
upper and lower branches of the integration
contour, where $x_C(t)$
is equal to
$x_+(t)$ and
$x_-(t)$, respectively, whereas the operator  $\hat\eta_{C}$
is equal to  $\hat\eta_t$ on both branches.
Introducing the symbol $\hat  T^{-1}$ to denote anti-time
ordering, Eq.~(\ref{3.4a}) takes the more explicit form
\begin{equation}
F[x_-,x_+]={\rm Tr}\left[\hat\rho_0^B
  \hat T^{-1} \exp\left( -\frac{i}{\hbar}
\int\limits^{t}_{0} dt' x_-\hat \eta
       \right)\hat {T}\  \exp\left( \frac{i}{\hbar} \int\limits^{t}_{0}
        dt' x_+ \hat\eta   \right)
\right]\ .
\label{3.4}\end{equation}
Assuming the explicit noise representation (\ref{noise}) through
the heat bath operators we observe that
the matrix element $\langle\bar{X}|\hat T\exp(i\int_0^t dt'x_+
\hat\eta/\hbar)|X\rangle$ and
the corresponding matrix element of the anti-time
ordered exponential in (\ref{3.4})
are given by the path integrals over $X_+$ and $X_-$
in (\ref{14}), respectively, whereas the trace in (\ref{3.4})
corresponds to the integrals over $X,\bar{X}$ and $\bar{X}'$
in (\ref{14}).

We now make the key
observation that will allow us to derive the quantum Langevin equation
(\ref{ql}):
The fluctuation part of the influence functional, defined
in
Eq.~(\ref{infl}),   is obtained from (\ref{3.4})
by setting $x=0$. Then both exponentials in (\ref{3.4})
carry the same argument $ ({i}/2{\hbar}) \int^{t}_{0}
        dt' y \hat\eta $ .
When expanding the product of the two exponentials
in a power series and performing the proper
time orderings, the
bath average
can be rewritten
as
\begin{equation}
F_{\rm fl}[y]=F[y/2,-y/2]=
 {\rm Tr} ~\left[\hat T\exp\left(\frac {i}{\hbar}\int\limits_{0}^{t}dt'
y_t\ \ad\hat{\eta}_t\right)\hat\rho^B_0\right] ,
\label{3.11}
\end{equation}
where $ \ad\hat{\eta}_t$ is the
adjoining operator associated with the noise operator
$\hat  \eta _t$.
This is defined as follows. When acting upon  an arbitrary operator
 $\hat{{\cal O}}$, the adjoining operator
is equal to half the anticommutator:
\begin{equation}
\ad\hat{\eta}_t\,\hat{{\cal O}}
\equiv [\hat{\eta}_t,\hat{{\cal O}}]_+/2.
\label{def}\end{equation}
The bath average (\ref{3.11})
corresponds to defining a noise average
of an arbitrary functional ${\cal O}[{\rm ad}\hat\eta]$ by
\begin{equation}
\langle {\cal O}[{\rm ad\ }\hat\eta] \rangle_{\hat\eta}^{\rm ad}
\equiv {\rm Tr}~{\cal O}[{\rm ad}\hat\eta] \hat{\rho}^B_0\ .
\label{aver}
\end{equation}
Note that
the adjoining operator
${\rm ad}\,\hat\eta $ acts also upon $\hat{\rho}^B_0 $.
The ket superscript
${\rm ad}$ emphasizes this
fact which is in contrast to
averages with respect to the ordinary noise operator
defined as in
(\ref{aver0}).
For example, the correlation function
is  calculated as
\begin{equation}
\langle {\rm ad\ }\hat\eta_t{\rm ad\ }\hat\eta_{t'}
\rangle_{\hat{\eta}}^{\rm ad}=
\frac 14{\rm Tr\ }[\hat{\eta}_t,[\hat{\eta}_{t'},\hat{\rho}^B_0]_+]_+ =
\frac 12{\rm Tr\ }\hat{\rho}_0^B[\hat{\eta}_t,\hat{\eta}_{t'}]_+ =
K_T(t,t')\ .
\label{cor}
\end{equation}

The adjoining operator  $\ad\hat  \eta _t$
has an important property crucial
to the further development:
Although the original noise
operator   $\hat  \eta _t$
has a nontrivial commutator (\ref{qn1})
with itself at a different time,
two adjoining operators
$\ad\hat  \eta _t$
and $\ad\hat  \eta _{t'}$
commute with each other, so that
they can be treated  as $c$-numbers.
Indeed, for an arbitrary operator $\hat{{\cal O}}$ we
verify that
\begin{equation}
[\ad\hat{\eta}_t,\ \,\ad\hat{\eta}_{t'}]\hat{{\cal O}} =
[\hat{\eta}_t,\hat{\eta}_{t'}]\hat{{\cal O}} + \hat{{\cal O}}[\hat{\eta}_{t'}
\hat{\eta}_t] = 0.
\label{ncom}\end{equation}
In all such calculations, we
may treat the quantum  noise $\ad\hat{\eta}_t$ as if it were a classical
noise $\eta_t$.

Furthermore, it is clear from the action (\ref{bath1})
that
${\rm ad\ }\hat\eta_t$ is a Gaussian noise variable, just as the classical
$ \eta _t$.
The correlation functions of any number of
operators  $\ad\hat{\eta}_t$
evaluated in the average
 (\ref{aver})
 are completely specified by their two-point
function (\ref{cor}) following Wick's expansion rule.
Since the
expectation values of the products of two $\eta_t$ or
of two
$\ad\hat{\eta}_t$ are identical,
$<\eta_t\eta_{t'}>_\eta \equiv
<\ad\hat{\eta}_t\ad\hat{\eta}_{t'}>_{\hat{\eta}}$,
all correlation functions must be identical as well---quantum
and classical noises are completely equivalent.

Substituting
(\ref{3.11}) into
(\ref{2.17})
and repeating the arguments leading to
Kubo's equation in Section III,
we again end up with a modification of
Kubo's stochastic Liouville equation, in which
the noise is operator-valued,
i.e., $\eta_t$ in (\ref{sla2}) and (\ref{3.3}) is
replaced by $\ad\hat{\eta}_t$.

Having shown how the operator-valued quantum noise turns into
a classical noise in Kubo's stochastic Liouville,
we are ready to
derive the
quantum Langevin equation (\ref{ql}).

\section{Quantum Langevin equation}

As the first step, we derive the evolution equation for
the system's density matrix operator
$\hat{\rho}^S_t$
whose
matrix elements are defined by (\ref{2.8}).
Its operator representation is
$\hat{\rho}^S_t ={\rm Tr}_B\hat{\rho}_t$
where $ {\rm Tr}_B$ denotes the
trace over bath degrees of freedom. Returning in Kubo's equation
(\ref{sla2}) to the
initial forward--backword variables $x_\pm$
of (\ref{newv}), we find the evolution  equation  for the
noisy
matrix elements associated with (\ref{2.8}):
\begin{equation}
i\hbar\partial_t\rho_t^S(x_+,x_-) =
\left[\hat{H}^S_+-\hat{H}^S_- +\frac{\gamma}{2m}(\hat x_+ -\hat x_-)
(\hat p_+ - \hat p_-) - (\hat x_+ -\hat x_-){\rm ad\ }\hat \eta_t\right]
\rho_t^S(x_+,x_-)\ ,
\label{kubo2} \end{equation}
where $\hat p_\pm \equiv -i\hbar\partial/\partial x_\pm$,
 and
$\hat{H}_\pm^S = H^S(\hat p_\pm,\hat x_\pm)$ is the system
Hamiltonian expressed in terms of the forward--backward
phase space
variables.

Of course, this equation  can also be obtained directly from
Eq.~(\ref{2.3}).
\comment{the Hamiltonian form of the forward--backward path integral (\ref{2.7}),
meaning that each of the path integrals in (\ref{2.7}) is to be
represented in the Hamiltonian form, just as we have derived
the Kubo equation (\ref{sla2}) from (\ref{2.17}).}
The Hamilton operators in the two exponentials
yield
directly
the difference $\hat H_+^S-\hat H_-^S$ in
(\ref{kubo2}), whereas the other two terms result from the
dissipation and fluctuation parts of the influence functional
in the path integral (\ref{2.7}).

Let $|x\rangle$ and $|p\rangle$ denote
eigenvectors of the canonically
conjugate system operators $\hat x$ and $\hat p$, respectively. Then
$\rho_t^S(x_+,x_-) =<x_+|\hat{\rho}_t^S|x_->$. We have the following
identies
\begin{eqnarray}
(\hat H_+^S-\hat H_-^S)\langle x_+|\hat{\rho}_t^S|x_-\rangle &= &
\langle x_+|\ [\hat H^S,\hat{\rho}_t^S]\ |x_-\rangle\ ;\label{id1}\\
(\hat x_+\hat p_+ +\hat x_-\hat p_-)
\langle x_+|\hat{\rho}_t^S|x_-\rangle &=&
\langle x_+|\ (\hat x\hat p\hat{\rho}_t^S-
\hat{\rho}_t^S\hat p\hat x)\ |x_-\rangle\ ;\\
(\hat x_-\hat p_+ +\hat x_+\hat p_-)
\langle x_+|\hat{\rho}_t^S|x_-\rangle &=&
\langle x_+|\ (\hat p\hat{\rho}_t^S\hat x - \hat x
\hat{\rho}_t^S\hat p)|x_- \rangle\ ,  \label{id3}
\end{eqnarray}
which can be proved by inserting suitable resolutions of unity,
$\int dp |p\rangle\langle p| = \int dx |x\rangle\langle x| = 1$,
between the operators on the right-hand sides.
Substituting Eqs.~(\ref{id1})--(\ref{id3}) into (\ref{kubo2}), we find the evolution equation
for the operator $\hat{\rho}_t^S[{\rm ad\ }\hat{\eta}]$
\begin{equation}
i\hbar\partial_t\hat{\rho}_t^S =[\hat{H}^S,\hat{\rho}_t^S] +
\frac{1}{2}[\hat{x},[\gamma\hat{p}/m-\hat{\eta}_t,\hat{\rho}_t^S]_+]
\ .
\label{kubo3}
\end{equation}
The average (\ref{aver}) of a solution to (\ref{kubo3}) gives
the density matrix operator of the system
\begin{equation}
\hat\rho_t^S =\langle\hat\rho_t^S[{\rm ad\ }\hat{\eta}]\rangle_{\hat\eta}
={\rm Tr}_B\hat\rho_t^S[{\rm ad\ }\hat{\eta}]\hat\rho_0^B
\label{49}\end{equation}

Now we turn directly to a derivation of equations of motion for system
operators. Let us recall first that in quantum mechanics there are two
equivalent representations of the mean-value $\langle\hat{\cal O}
\rangle_t$ of any dynamical variable
$\hat{\cal O}$ (with or without explicit time dependence)
at a time $t$
\begin{equation}
\langle\hat{\cal O}\rangle_t = {\rm Tr}\hat{\cal O}\hat\rho_t =
{\rm Tr}\hat{\cal O}_t\hat\rho_0 \ ,
\label{55}
\end{equation}
where $\hat\rho^S_t$ is the density matrix operator at the time $t$,
and $\hat{\cal O}_t$ is the Heisenberg operator
coinciding with
$\hat{\cal O}$ at $t=0$. The first equality expresses the mean-value in the
Schr\"odinger picture, while the second one determines it in the
Heisenberg representation. Equation (\ref{55}) can also be regarded
as a definition of the Heisenberg operator $\hat{\cal O}_t$. The
consistency of this definition is guaranteed by
the time independence
of the commutation relations $[\hat{\cal O}, \hat{\cal O}'] =
[\hat{\cal O}_t, \hat{\cal O}_t'] $ for any two operators.
This, in turn,
is a consequence of
the unitarity of the temporal evolution of the density
matrix described by the Neumann equation (\ref{neq}).

The evolution of a quantum systems with dissipation is determined by
a generalization of the Neumann equation (\ref{kubo3}).
We shall demonstrate in analogy with (\ref{55}),
there exists a consistent
definition
of noisy
Heisenberg operators,
if the noise obeys
the commutation relation
(\ref{qn1}).
For any product of system operators $\hat{\cal O}^S =
\prod_{k=1}^n \hat{\cal O}^{(k)}$,
a product of noisy Heisenberg system operators
is defined by the equality
\begin{equation}
{\rm Tr}_S \hat{\cal O}^S\hat\rho_t^S[{\rm ad}\hat\eta]=
{\rm Tr}_S\prod_{k=1}^n \hat{\cal O}^{(k)}\hat\rho_t^S[{\rm ad}\hat\eta]
\equiv
{\rm Tr}_S\prod_{k=1}^n \hat{\cal O}^{(k)}_t[\hat\eta]\hat\rho_0^S =
{\rm Tr}_S \hat{\cal O}^S_t[\hat\eta]\hat\rho_0\ ,
\label{56}
\end{equation}
where the noisy density matrix operator satisfies Kubo's
stochastic equation (\ref{kubo3}), and
${\rm Tr}_S$ implies a trace over system degrees of freedom.
One can think of the relation (\ref{56}) as a mapping of an algebra
of system operators at the
time $t=0$ onto their algebra at a time
$t$. Taking a trace of the evolution equation (\ref{kubo3}), we
see that the unit operator $\hat I$
remains invariant under the mapping (\ref{56}), i.e.,
$\dot{\hat I}_t[\hat\eta] = 0$, so that ${\hat I}_t[\hat\eta] \equiv \hat I$.
The consitency of the definition (\ref{56})
is guaranteed by
the time independence of
the commutation relations
\begin{equation}
[\hat{\cal O}^S,\hat{\cal O}'^S]=
[\hat{\cal O}^S_t[\hat\eta],
\hat{\cal O}'^S_t[\hat\eta]]\ ,
\label{ocom}
\end{equation}
for any two $\hat{\cal O}^S_t[\hat\eta] $ and $\hat{\cal O}'^S_t[\hat\eta]$.
At  $t=0$,  equation (\ref{ocom})
is trivially fulfilled.
To prove the time independence,
we derive the noisy Heisenberg
equation for $\hat{\cal O}^S_t$.
Differentiating (\ref{56}) with respect to time and making use of both
(\ref{kubo3})  and (\ref{56}), we obtain
\begin{equation}
i\hbar\dot{\hat{\cal O}}_t{}^S
=[\hat{\cal O}^S_t, \hat H^S_t] +\frac 12\left[\gamma
\hat p_t/m -\hat\eta_t,\ [\hat{\cal O}^S_t, \hat x_t]\right]_+\ ,
\label{58}
\end{equation}
where $\hat H^S_t=H^S(\hat p_t, \hat x_t) $.
For a noise operator satisfying the commutation rule (\ref{qn1}),
this equation leads indeed to
(\ref{ocom}) (see \cite{gardiner}). Equivalently,
we
demonstrate that Eq.~(\ref{58})
coincides
with the Heisenberg equations of motion of the
entire system (the particle in
a heat bath), comprising
the quantum Langevin equation
(\ref{ql}).

For an arbitrary system operator we have
\begin{eqnarray}
i\hbar\dot{\hat{\cal O}}_t{}^S &=& [\hat{\cal O}_t^S,\hat{H}_t] =
[\hat{\cal O}_t^S, \hat H^S_t] + \frac 12
\left(\frac{2\gamma}{\pi}\right)^{1/2}
\left[\int\limits_{0}^{\infty}d\omega\dot{\hat{X}}_{\omega t} ,
[\hat{\cal O}_t^S, \hat x_t]\right]_+\
;\label{sys}\\
i\hbar\dot{\hat{X}}_{\omega t} &=&[\hat{X}_{\omega t} ,\hat{H}_t]\  ,\ \ \ \
i\hbar\dot{\hat{P}}_{\omega t} =[\hat{P}_{\omega t} ,\hat{H}_t]\ ,
\label{oss}
\end{eqnarray}
where $\hat H_t$ is the canonical Hamilton operator corresponding
to the total action (\ref{act}) defined by (\ref{sysac}), (\ref{bath1})
and (\ref{bath3}):
\begin{equation}
\hat{H}_t=\hat{H}^S_t+\hat{H}^B_t+\hat{H}^I_t=
\hat{H}^S_t +\frac 12\int\limits_{0}^{\infty}d\omega
\left\{\left[\hat{P}_{\omega t} -\left(\frac{2\gamma}{\pi}\right)^{1/2}
\hat{x}_t\right]^2  +\omega^2\hat{X}_{\omega t}^2\right\} .
\label{th}\end{equation}
The $ \omega $-integral in this Hamiltonian
produces a linearly divergent term $\propto \hat x^2/2$,
the "frequency shift term" discussed at length in Ref.~\cite{leggett}.
The divergence is
due to the specific form of the interaction (\ref{bath3})
which generates the desired
time-independent friction coefficient
$\gamma$ in the Langevin equation \cite{zwanzig}.
The same divergence
appears also
in the Hamiltonian form of the forward-backward path integral (\ref{2.7}).
The divergence
is canceled  by an equal divergence
in the dissipation part of
the influence functional arising from the coupling to momenta of heat
bath oscillators (the term $\sim xP_\omega$ in (\ref{th})) \cite{leggett},
p.602. For this reason,
Kubo's stochastic equation has only finite terms.
This pleasant cancelation is absent
if
the system is coupled directly to
the
positions of the bath oscillators
 rather than to their velocities,
as in (\ref{bath3}).
In that case, the initial potential $V(x)$ requires a
divergent counter term in order to obtain a
finite
Kubo equation.
The entire issue is, of course, somewhat academic since it is
a consequence of insisting upon
a constant friction, which is physical only at low frequencies,
much lower than the collision rate in the system.
For larger frequencies,
the physical behavior is described by the
Drude friction $\gamma(\omega) \propto 1/(1+ \omega ^2/ \omega _D^2)$,
and there is no divergence at all
\cite{zwanzig}, \cite{kleinert3}.

Imagining the presence of a frequency cutoff in (\ref{th}),
we solve the oscillator
Heisenberg equations (\ref{oss}) and find
\begin{equation}
\left(\frac{2\gamma}{\pi}\right)^{1/2}
\int\limits_{0}^{\infty}d\omega\dot{\hat{X}}_{\omega t} =
\gamma\dot{\hat x}_t -\hat\eta_t = \gamma\hat p_t/m
-\hat \eta_t \ ,
\label{xdot}
\end{equation}
where we have used the representation (\ref{3.6}) for the noise operator.
Substituting (\ref{xdot}) into (\ref{sys})
and comparing it with (\ref{58}) we conclude that
\begin{equation}
\hat{\cal O}_t^S[\hat\eta] =
e^{i\hat Ht/\hbar}\hat{\cal O}^Se^{-i\hat Ht/\hbar} = \hat{\cal O}^S_t\ .
\label{hso}
\end{equation}
Thus,
the solutions of the evolution equation (\ref{58})
coincide with
 Heisenberg system operators. Hence, the temporal evolution
 is unitary and (\ref{ocom}) is satisfied. Note that the
system Hamiltonian is no longer an integral of motion,  $
 d\hat H^S_t/dt
\neq 0$, in contrast with the Hamiltonian driving
the ordinary Heisenberg
equations.

Setting $\hat{\cal O}^S_t$ equal to $\hat x_t$ and $\hat p_t$ in
Eq.~(\ref{58}) or, equivalently, in
Eq.~(\ref{sys}), we obtain the Heisenberg
equations of motion:
\begin{eqnarray}
\dot{\hat{x}}_t &= &\hat{p}_t/m\ ,\label{ql1}\\
\dot{\hat{p}}_t &= &-V'(\hat{x}_t) -\gamma\hat{p}_t/m +\hat{\eta}_t
\label{ql2},
\end{eqnarray}
which are equivalent to (\ref{ql})

Equation (\ref{hso}) ensures the
unitarity of the temporal evolution and the time independence
of the canonical
commutation relation $[\hat x_t,\hat p_t]=i\hbar$.
This guarantees also that correlation functions of noisy
operators within the $\hat  \eta $ average agree with the
ordinary quantum-mechanical correlation functions of the
system described by the Hamilton operator (\ref{th}):
\begin{equation}
{\rm Tr}_S\hat\rho_0^S\langle[\hat{\cal O}_t^S[\hat\eta],
\hat{\cal O}'^S_{t'}[\hat\eta]]\rangle_{\hat\eta} =
{\rm Tr\ }\hat\rho_0 [\hat{\cal O}^S_t,\hat{\cal O}'^S_{t'}]\ .
\label{corf}
\end{equation}

The
property (\ref{qn1})
of the quantum noise is crucial for the unitarity of the time
evolution of the Heisenberg operators.
To see what happens if the noise were commutative, we replace
${\rm ad\ }\hat\eta$ in (\ref{kubo3})
by a $c$-valued Gaussian noise
$\eta$, which we  are apparently allowed to do
after the above observations on the equivalence of averages.
Then we can again define the noisy Heisenberg representation
by (\ref{56}) and derive (\ref{58}) in a similar way.
However, we can never prove the consistency of (\ref{58})  and (\ref{56});
Eqs. (\ref{ocom}) and (\ref{hso}) are no longer valid.
For a commutative noise,
the violation of the canonical commutation relation as time proceeds
is seen from (\ref{ql1}), (\ref{ql2}), most simply for $V=0$.
This break-down of the unitarity in the temporal evolution
would prevent us from identifying (\ref{58}) with the quantum
Langevin equation.

Thus we have shown that the Feynman-Vernon path integral
description of quantum
Brownian motion is completely equivalent to a description
in terms
of
the quantum Langevin equation (\ref{ql}).
Both  descriptions
can be used on equal footing
to study dissipative processes
in quantum mechanics.


\begin{references}

\bibitem{senitzky} J.R. Senitzky, {\em Phys.Rev.} {\bf 119} (1960) 670;
{\bf 124} (1961) 642.
\bibitem{lax} M. Lax, {\em Phys.Rev.} {\bf 129} (1963) 2324.
%
\bibitem{gardiner}C.W. Gardiner, {\em IBM J.Res.Develop.} {\bf 32}
(1988) 127.
  %
 \bibitem{ullersma} P. Ullersma, {\em Physica} {\bf 32} (1966) 27.
\bibitem{kac} R. Benguria and M. Kac, {\em Phys.Rev.Lett.} {\bf 46}
(1081) 1.
\bibitem{mazur} G.W. Ford, M. Kac and P. Mazur, {\em J.Math.Phys.}
{\bf 6} (1965) 504.
\bibitem{zwanzig} R. Zwanzig, {\em J.Stat.Phys.} {\bf 9} (1973)215.
\bibitem{ford} G.W. Ford and M. Kac, {\em J.Stat.Phys.} {\bf 46}
(1987) 803.
%
\bibitem{kubo} R.~Kubo, {\em J.Math.Phys.} {\bf 4} (1963) 174;\\
R.~Kubo, M.~Toda and N.~Nashitsume,
    {\em Statistical Physics II (Nonequilibrium Statistical Mechanics)},
    Springer-Verlag, Berlin, 1985 (chp.~2).
%
\bibitem{leggett} A.O. Caldeira and A.J. Leggett, {\em Physica}
{\bf 121A} (1983) 587.

\bibitem{fey} R. Feynman and A. Hibbs, {\em Quantum Mechanics and
Path Integrals} (McGraw-Hill, New York, 1965).

\bibitem{vernon} R. Feynman and F.L. Vernon, {\em Annals of Physics}
{\bf 24} (1963) 118.
\bibitem{kleinert} H. Kleinert, {\em Path Integrals in Quantum
Mechanics, Statistics and Polymer Physics}, sec. edition,  World Scientific,
Singapore, 1994.

\bibitem{schmid}
{A.~Schmid}, {\em J.~Stat.\ Phys.} {\bf 59} (1990) 855.
\bibitem{qcle}
A.~Schmid, {\em J.~Low Temp.\ Phys.} {\bf 49\/} (1982) 609;\\
{U.~Eckern}, {W.~Lehr}, {A.~Menzel-Dorwarth}, {F.~Pelzer},
and A.Schmid,
{\em J.Stat.Phys.} {\bf 59} (1990) 855.
\bibitem{cite1}
	 See Section 18.6 in
Ref.~\cite{kleinert}
\bibitem{cite2}
	 See Eq.~(18.162) in Ref.~\cite{kleinert}
\bibitem{kleinert3}
	 See Eq.~(3.156) in Ref.~\cite{kleinert}

\end{references}
\end{document}